\documentclass{article} %
\usepackage[submission]{colm2025_conference}

\usepackage{graphicx} %
\usepackage{natbib}  %
\usepackage{caption} %
\usepackage{algorithm}
\usepackage{listings}
\usepackage{algorithmic}
\usepackage{amsmath}

\usepackage{multirow}
\usepackage[outdir=./]{epstopdf}
\usepackage{enumitem}
\usepackage{caption}
\usepackage{subcaption}
\usepackage{subfloat}
\usepackage{newfloat}
\usepackage{graphicx}
\usepackage{svg} %
\usepackage[normalem]{ulem}
\usepackage{framed}
\usepackage{mdframed}
\usepackage{xcolor}
\usepackage{lipsum}
\usepackage{float}

\usepackage{CJKutf8}
\usepackage{tabularx}
\usepackage{microtype}
\usepackage{hyperref}
\usepackage{url}
\usepackage{booktabs}
\usepackage{tikz}
\usepackage{xcolor} %
\usepackage{siunitx}

\definecolor{darkblue}{rgb}{0, 0, 0.5}
\hypersetup{colorlinks=true, citecolor=darkblue, linkcolor=darkblue, urlcolor=darkblue}

\title{An Evaluation of Cultural Value Alignment in LLM}

\author{Nicholas Sukiennik \\
Department of Electronic Engineering\\
Tsinghua University\\
Beijing, China \\
\texttt{sukiennikn10@mails.tsinghua.edu.cn}
\And %
Chen Gao \\
BNRist\\
Tsinghua University\\
Beijing, China\\
\texttt{chgao96@gmail.com}
\And
Fengli Xu \\
Department of Electronic Engineering\\
Tsinghua University\\
Beijing, China \\
\texttt{fenglixu@tsinghua.edu.cn}
\And
Yong Li \\
Department of Electronic Engineering\\
Tsinghua University\\
Beijing, China \\
\texttt{liyong07@tsinghua.edu.cn}
}

\begin{document}

\ifcolmsubmission
\fi

\maketitle

\begin{abstract}
LLMs as intelligent agents are being increasingly applied in scenarios where human interactions are involved, leading to a critical concern about whether LLMs are faithful to the variations in culture across regions. Several works have investigated this question in various ways, finding that there are biases present in the cultural representations of LLM outputs. To gain a more comprehensive view, in this work, we conduct the first large-scale evaluation of LLM culture assessing 20 countries' cultures and languages across ten LLMs. With a renowned cultural values questionnaire and by carefully analyzing LLM output with human ground truth scores, we thoroughly study LLMs' cultural alignment across countries and among individual models. Our findings show that the output over all models represents a moderate cultural middle ground. Given the overall skew, we propose an alignment metric, revealing that the United States is the best-aligned country and GLM-4 has the best ability to align to cultural values. Deeper investigation sheds light on the influence of model origin, prompt language, and value dimensions on cultural output. Specifically, models, regardless of where they originate, align better with the US than they do with China. The conclusions provide insight to how LLMs can be better aligned to various cultures as well as provoke further discussion of the potential for LLMs to propagate cultural bias and the need for more culturally adaptable models. 

\end{abstract}

\section{Introduction}
Large language models (LLMs), with their astonishing linguistic capabilities, demonstrate human-like reasoning and behavioral abilities across a multitude of tasks~\citep{zhao2023survey}, and with their widespread usage, they have more and more obligated to provide realistic and accurate responses. That is, alignment in LLMs is essential for these models to provide trustworthy responses to users. Specifically, alignment involves ``ensuring that AI systems pursue goals that match human values or interests rather than unintended and undesirable goals''~\citep{pan2022effects}. Therefore, in this work, we present a comprehensive evaluation of cultural value alignment across a gamut of LLMs, with input prompts that cover 20 countries and their respective languages. Our primary aim is to determine which models align best to cultural values overall, and which countries are best \textit{aligned to}. We further examine the role of model origin on alignment results. To evaluate culture, we adopt a well-known survey known as the Values Survey Module \citep{hofstede_culture_1980}, a six dimension, 24-item survey where aggregated human responses form the ground truth culture scores for each dimension and country. While prior works investigate culture bias in LLMs, ours is the first to systematically address it on a broad, comprehensive scale over a range of models, countries, and languages. We also present extensive factor analysis to understand underlying causes and correlations, such as the influence of prompt language, model origin, and external factors.

There are several research efforts about understanding and quantifying the presence of alignment in LLM \citep{ngo_alignment_2024,ryan_unintended_2024,cao_assessing_2023}. By discovering the presence of cultural misalignment on a large scale, our work is also related to these works on LLM bias. Bias is a particular form of misalignment that occurs when the model reflects only the perspectives of a certain sub-group or population \citep{li2022herb}, which can serve to reinforce asymmetries among groups within a society \citep{bender2021dangers}.  
The presence of bias in LLMs is an unsolved problem due to the vast scope and scale of data that is used to train them \citep{johnson_ghost_2022}. More specifically, culture bias is of particular interest, especially for applications of LLMs that require them to interface with individuals from different parts of the world, such as customer service chatbots, etc. 
Several works confirm the presence of a Western cultural bias in LLMs due largely to the sources of the corpora involved in their training \citep{naous_having_2024, arora_probing_2023, navigli_biases_2023}.
Other works address forms of bias that are unrelated to culture, such as ideologies \citep{buyl_large_2024, santurkar_whose_2023}, social identity biases \citep{hu_generative_2024}, whereas \citep{yin2025safeworld} address cultural and legal appropriateness in social interactions.

We briefly describe the works that are similar to ours and what we do differently to contribute towards a substantial increase in understanding of LLMs' cultural alignment. The work that is most similar to ours is BLEnD \citep{myung_blend_2024}, which tests the ability of LLMs to adhere to culturally specific knowledge. Their work evaluates LLMs' performance using several models and languages to determine the presence of region bias and language bias, discovering that the results heavily favor highly represented languages (\textit{e.g.} English and Spanish), as opposed to low-resource countries such as Nigeria and Ethiopia. However, their work does not address the values component of culture. In contrast, a handful of works address cultural values, but only in a limited context. 
Namely, \citet{alkhamissi_investigating_2024} evaluate the cultural alignment of four LLMs on their cultural alignment for two cultures, Egypt and the United States, using the languages English and Arabic, using Hofstede's survey to assess performance. \citet{cao_assessing_2023} evaluate cultural values alignment using five languages and country-roles and only one GPT model, similarly using the Hofstede survey. Finally, \citet{tao_auditing_2023} evaluate 107 countries' cultures using only English prompts, and three GPT models (GPT-3, GPT-3.5-Turbo, GPT -4), using the Integrated Values Survey to measure culture \citep{inglehart2005christian}. Each of these works paints a small portion of an overall picture of cultural values alignment in LLMs, but due to the scarcity of models, countries, and languages tested, as well as the inconsistency between methodologies across works, such picture is far from complete. Inspired by their methodologies and with the goal of completing this picture, our work scales up the analysis using 10 models, 20 countries and languages, and conducts in-depth analysis to understand the underlying mechanisms and correlations of cultural bias.

In addition to investigating the influence of prompt language on cultural alignment, we are the first work to address the concern of models propagating cultural values based on their origin. The overall contributions of this work are as follows:

\begin{enumerate}
    \item We propose to study the overall state of LLM alignment on countries, and rank the top-performing models and countries using a proposed alignment metric. %
    \item We conduct large-scale analysis on the results to understand the influence of model origin, language, and model size on alignment, among other factors.   %
    \item We find that the United States is the most closely aligned country across all models by a wide margin, and that GLM-4 performs best on cultural alignment despite its small size.   %
\end{enumerate}

\section{Methodology}
In order to evaluate the overall state of cultural alignment of LLMs, we approach the problem from two angles: 1) which LLMs align best across cultures, and 2) which countries are aligned to the best. In answering these two questions, we can determine both the ability and extent of different LLMs to embody a realistic cultural value system while also determining the presence of bias in LLMs as a whole. Our study also examines the relationship between alignment and prompt language, focusing on four types of language prompting results:
\begin{enumerate}[leftmargin=*]
    \item \textbf{Aligned Prompt Results}, meaning that the prompt language is aligned with the mainstream spoken or official language of a given country.
    \item \textbf{English} and \textbf{Chinese} prompt results, motivated by the fact that the models tested originate in the US (mainstream language: English) and China (official language: Chinese). Beyond prompt language, a deeper investigation of the implications of model origin is also presented in section \ref{sec:originanalysis}.
    \item \textbf{Language Average Results} which is the average of the results over all promoted languages. In other words, for each culture-identity, the LLM is prompted with all 20 languages, regardless of alignment. The purpose of this is to determine whether there is a convergence of culture for each country-role beyond the influence of prompt language.
\end{enumerate}

The cultural assessments of obtained by prompting each LLM with the questions of the Values Survey Module, a robust questionnaire by Geert Hofstede \citep{hofstede_culture_1980} that is seen as the gold standard of cultural studies. The survey consists of 24 questions that are broken down into six dimensions. The detailed use of the VSM is provided in Appendix Section \ref{sec:appendixhofstede}.

For this evaluation, we select 20 countries and their corresponding primary language, shown in Appendix Table \ref{tab:alphabetical_countries_languages}. The countries and languages are selected via a process of consideration based on several factors. Most importantly, we aim to choose countries whose language has a large number of native speakers while also being \textit{relatively} exclusive to that country.

\subsection{Alignment Measurement Metric}

In the course of our investigation, we discover the presence of a moderate "global average" culture to which all models tend to conform, regardless of country-role or prompt language, allowing us to better quantify cultural alignment given the presence of a bias-inducing factor. We address this by proposing an evaluation metric called the deviation ratio. The deviation ratio is the ratio between the deviation of an LLM's cultural representation from the global average culture and its difference from the ground truth, as seen in equation \ref{eqn:deviationratio}: 
\begin{equation}
    \text{Deviation Ratio} 
    =
    =\frac{\frac{1}
    {6}\sum_{d\in D} |\text{GT}_d - \overline{\text{GT}}_d|}{\frac{1}{n}\sum_{i=1}^n \text{Difference}_i},
    \label{eqn:deviationratio}
\end{equation}

\noindent where $D$ contains the six cultural dimensions, 
$\text{GT}_d$ is the ground truth value for dimension $d$, $\overline{\text{GT}}_d$ is the global average ground truth for dimension $d$, $\text{Difference}_i$ represents the individual model differences from ground truth, $n$ is the number of trials being averaged, in this case three. 
 Without this metric, the countries whose inherent culture falls close to the global average culture will automatically show more alignment, which is not a reflection of the LLM's capabilities. In contrast, if a country's culture is far from the global average, yet still closely aligned with the ground truth, then its evaluation will be much better than a country whose culture is closer to the global average and just as closely aligned with the ground truth, making for a more reliable evaluation.

\section{Experimental Setup}

\begin{table*}[htbp]
\centering
\footnotesize
\begin{CJK*}{UTF8}{gbsn}
\begin{tabular}{>{\raggedright\arraybackslash}p{0.07\textwidth}>{\raggedright\arraybackslash}p{0.15\textwidth}>{\raggedright\arraybackslash}p{0.50\textwidth}>{\raggedright\arraybackslash}p{0.15\textwidth}}
\toprule
\textbf{Lang} & \textbf{System Role} & \textbf{Prompt \& Response Options} & \textbf{Sample Response} \\
\midrule
EN & Your role is an average person from \{country\}.
& In choosing an ideal job, having a boss you can respect is: (1) of utmost importance; (2) very important; (3) of moderate importance; (4) of little importance; (5) of very little or no importance?
& "2 - very important" \\

\bottomrule
\end{tabular}
\end{CJK*}
\caption{Prompting mechanism with system role, survey question and response options, and a sample response.}
\vspace{-0.2cm}
\label{tab:language_features_prompting_examples}
\end{table*}

Ten LLMs are prompted with each of the 20 countries as a role and in each of the corresponding countries' languages, i.e. 400 times each. Five US-origin models and five China-origin models are selected, as to facilitate understanding of the influence of model origin on cultural alignment. The models are chosen to represent those that are most popularly used in their respective countries' origin as well as globally. The models and their respective information are provided in Table \ref{tab:llm_specs}.
The models were called using a temperature of zero as to reduce deterministic outputs and increase reproducibility. Furthermore, each country-language prompt was called three times and averaged for each model, further decreasing the potential for outliers or randomness. 

The specific prompting scheme is displayed in Table \ref{tab:language_features_prompting_examples}. The system role was also appended with a statement to "make only one choice and always include a numerical value in your response." The system role, including country names, was also translated to each respective language as to prevent the influence of system role language affecting output culture. Although the system role specifies to include a numerical response, some models offered explanations for why such response was chosen, whereas some did not. Either way, only the numerical response was extracted and used for analysis.

\begin{table}[h]
\centering
\small

\begin{tabular}{llrl}
\toprule
\textbf{Model} & \textbf{Company} & \textbf{Size (B)} \\
\midrule
GPT-3.5-Turbo~\citep{gpt35openai} & OpenAI &$ \sim20$ \\
GPT-4~\citep{gpt4openai} & OpenAI & $\sim1750$  \\
GPT-4o~\citep{hurst2024gpt} & OpenAI & $\sim200$  \\
Gemini-1.5~\citep{gemini} & Google & $\sim1500 $ \\
LLaMa-3~\citep{touvron2023llama} & Meta & 70 \\
Deepseek-v2.5~\citep{deepseekv2} & Deepseek & $236$ \\
GLM-4~\citep{glm2022} & Zhipu AI/Tsinghua & 9 \\
Qwen-2.5-7B-Instruct~\citep{qwen2.5} & Alibaba & 7  \\
Qwen-2.5-32B-Instruct & Alibaba & 32  \\
Qwen-2.5-72B-Instruct & Alibaba & 72  \\
\bottomrule

\end{tabular}
\caption{The large language models used in this study and their specifications ($\sim$ denotes approximation, as the parameters for certain models have not been publicly disclosed)}
\label{tab:llm_specs}
\end{table}

Outside of the overall evaluation of model and country alignment, we also delve into the regional implications of alignment. While the question of which countries are aligned to best for each LLM is worth inspecting, this work focuses on two special regional aspects that, given the geographical implications of LLM development, are most insightful: the comparison between China-origin and US-origin models. Finally, we delve into the influencing factors towards cultural alignment, including prompt language, model size, and external factors such as country GDP and digital population size per country. 

Through the above cultural evaluation method, we aim to answer the following research questions: 
\begin{itemize}
    \item \textbf{RQ1:} Which LLMs exhibit the best cultural alignment? 
    \item \textbf{RQ2:} Which country is aligned to the best across all LLMs? 
    \item \textbf{RQ3:} What is the influence of model origin and language on cultural alignment? 
    \item \textbf{RQ4:} What external factors could lead to cultural misalignment or bias? 
\end{itemize}

\section{Analysis Results}

\subsection{Results on Country and Model Basis}

\begin{figure}
    \centering
    \includegraphics[width=0.90\linewidth]{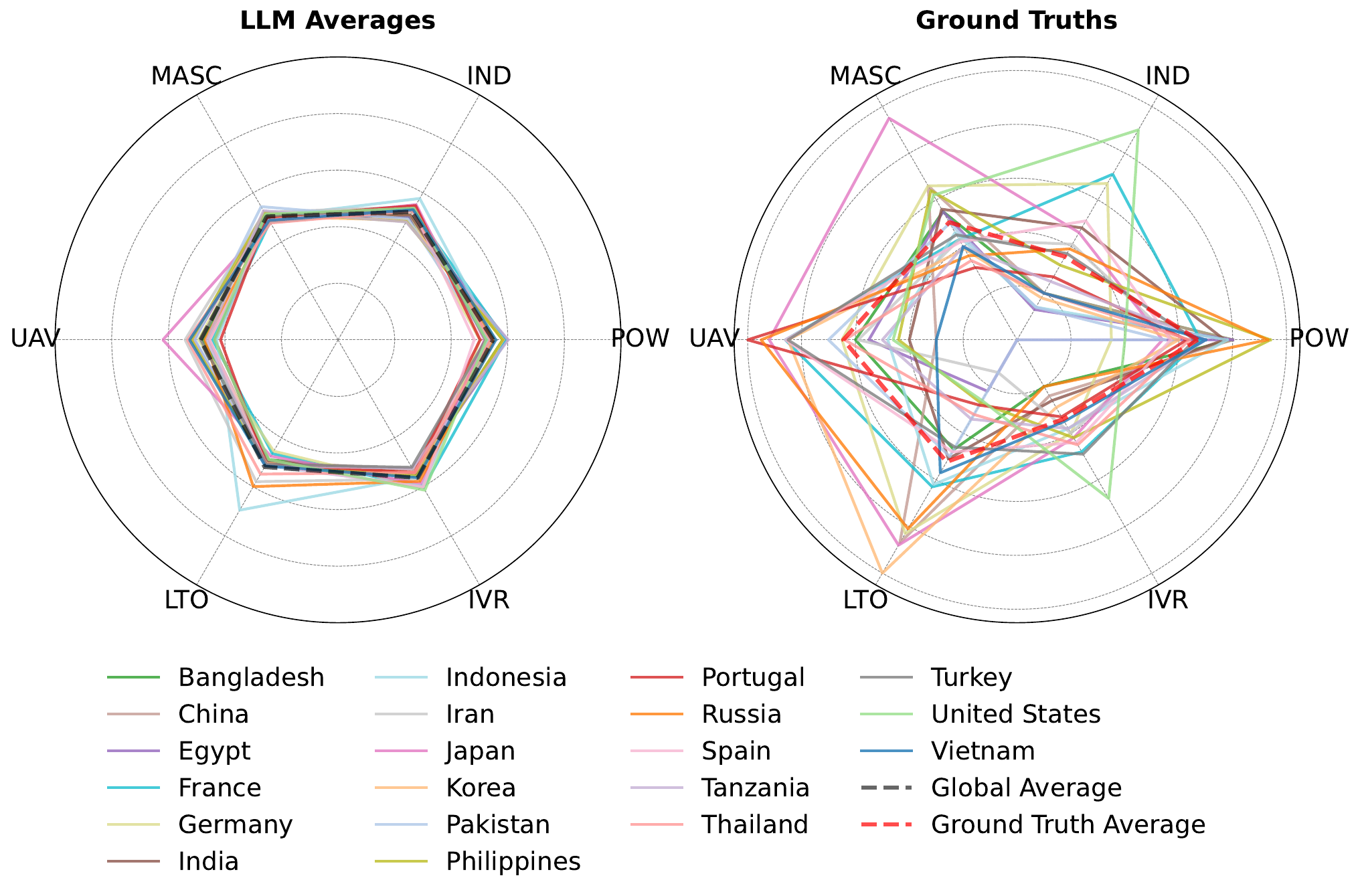}
    \caption{Comparison of ground truth and raw country results, with average ground truth and average of all LLM results.}
    \label{fig:raw_results_radar_averages}
\end{figure}

First, we evaluate the overall status of LLM alignment with countries' cultures in comparison with ground truth values over six dimensions~\footnote{The raw results of all models for each country can be found in Appendix Figure \ref{fig:country_raw_results_radars}}. In Figure \ref{fig:raw_results_radar_averages}(a), the raw results for each country are averaged over all models and plotted on a radar chart with 6 axes. The average over all countries is shown in black. This is shown alongside the ground truths of all countries and the average thereof in Figure \ref{fig:raw_results_radar_averages}(b). Immediately, we can make two observations: 1) the cultural representations output by LLMs for all countries are very close together, sitting close to the middle of the axis for nearly all dimensions, and 2) that the LLM cultural representations, as a whole, are markedly different from their ground truth counterparts.

\begin{figure}
\centering
\begin{minipage}{0.55\textwidth}
    \centering
    \includegraphics[width=\linewidth]{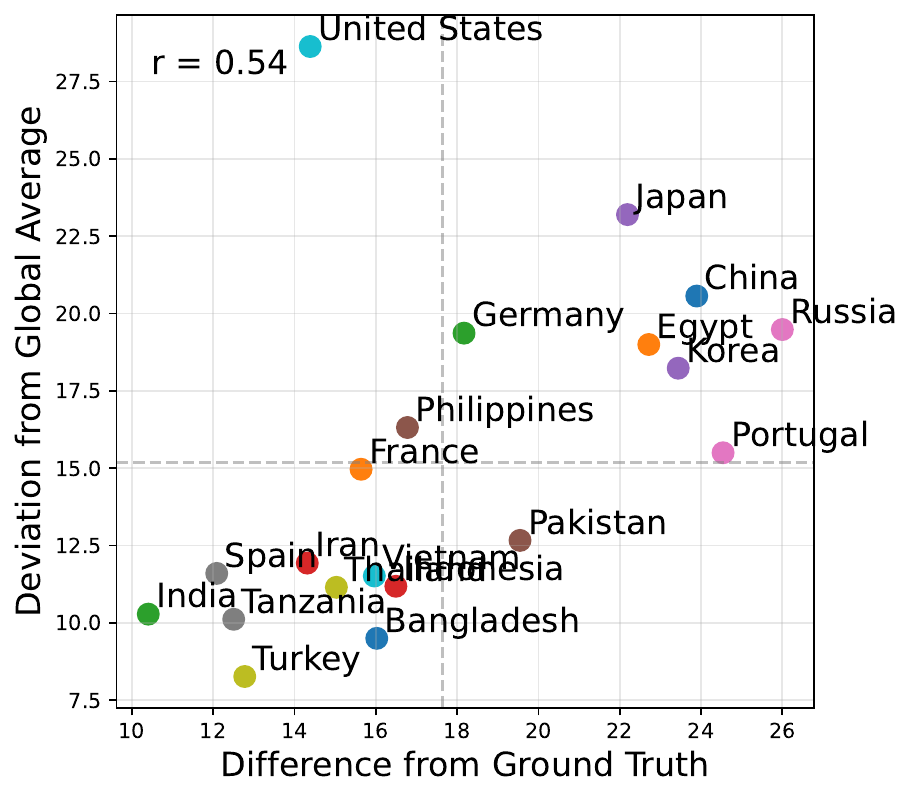}
    \captionof{figure}{Deviation from global average vs. difference from ground truth.}
    \label{fig:countries_scatter}
\end{minipage}
\hfill
\begin{minipage}{0.4\textwidth}
    \centering
    \footnotesize
    \begin{tabular}{l S[table-format=1.2]}
    \toprule
    \textbf{Country} & \textbf{Deviation Ratio} \\
    \midrule
    United States & 1.99 \\
    Germany & 1.13 \\
    Japan & 1.03 \\
    India & 1.02 \\
    Spain & 1.02 \\
    Philippines & 0.98 \\
    France & 0.96 \\
    China & 0.87 \\
    Iran & 0.83 \\
    Egypt & 0.80 \\
    Korea & 0.79 \\
    Russia & 0.77 \\
    Indonesia & 0.76 \\
    Thailand & 0.75 \\
    Vietnam & 0.73 \\
    Tanzania & 0.69 \\
    Pakistan & 0.67 \\
    Turkey & 0.66 \\
    Portugal & 0.62 \\
    Bangladesh & 0.59 \\
    \bottomrule
    \end{tabular}
    \captionof{table}{Country ranking.}
    \label{tab:deviation_ratios}
\end{minipage}
\end{figure}

To further examine the state of LLM cultural alignment on the country basis, the results for each country are broken down into two metrics: the difference from the ground truth and the deviation of the LLMs' output from the global average. In Figure \ref{fig:countries_scatter}, we can clearly see that, with the exception of the United States, there is a very strong linear relation between the two metrics. From this, we can conclude that evaluating alignment with respect to the ground truth would be insufficient. If only ground truth difference were considered, then India would be the most well-aligned country. But because India's culture sits very close to the global average, the ground truth difference being close is merely incidental and does not reflect LLM's ability to adapt. Therefore, the true alignment can be derived by looking at the ratio between the y and x axes, the results of which are seen in Table \ref{tab:deviation_ratios}.

The conclusions of this figure motivate the rest of the study to examine not purely cultural alignment, but\textit{ culture alignment given a tendency to converge on a global average}. In light of this aim, we devise Equation \ref{eqn:deviationratio}, which is essentially $y$ over $x$ in Figure \ref{fig:countries_scatter}, to achieve a more reliable metric for alignment.

\begin{figure}
    \centering
    \begin{tikzpicture}
        \node[anchor=south west, inner sep=0] (image) at (0,0) 
        {\includegraphics[width=1.00\linewidth]{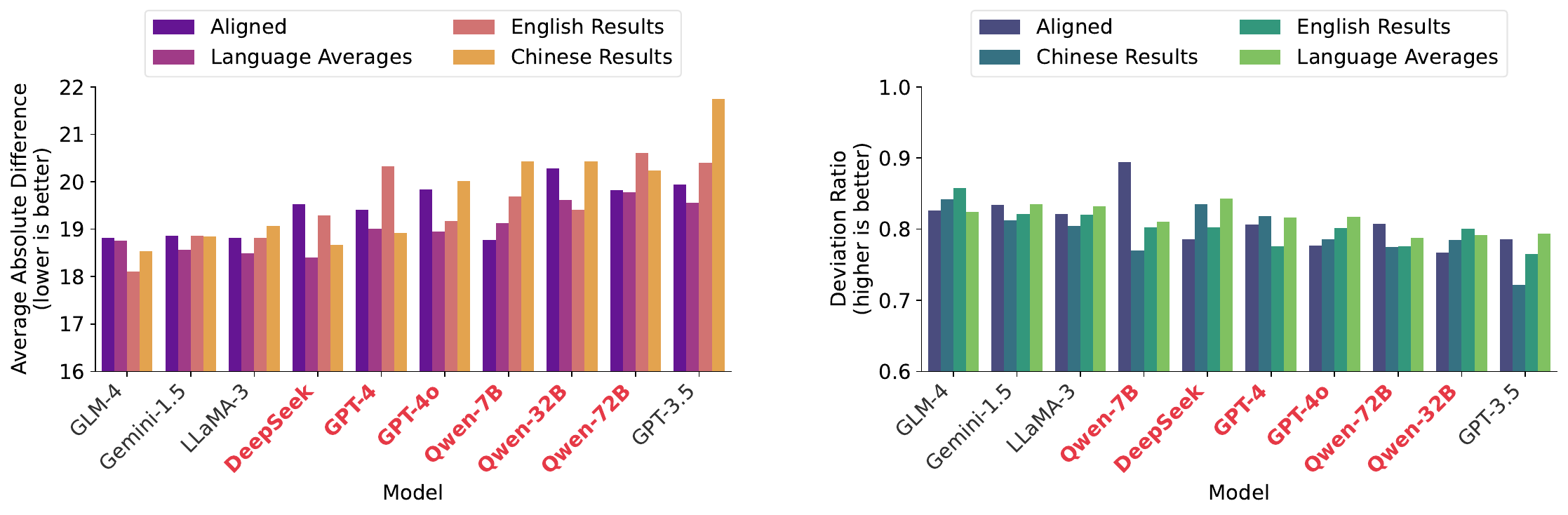}};
        \begin{scope}[x={(image.south east)}, y={(image.north west)}] %
            \node[anchor=north west, text=black] at (0.02, 0.98) {\textbf{(a)}};
            \node[anchor=north west, text=black] at (0.52, 0.98) {\textbf{(b)}};
        \end{scope}
    \end{tikzpicture}
    \caption{A comparison of model evaluation using two metrics: difference from ground truth (lower = better), and deviation ratio (higher = better). Each model ranking contains results with four prompting methods. The models that differ in rank between the two figures are highlighted in red.}
    \label{fig:avg-differences}
\end{figure}

We then evaluate the overall performance of all models using the new metric and the original difference from ground truth, as seen in Figures \ref{fig:avg-differences}(a) and (b). Although both figures rank the first three models in the same position, the following six are in a distinct order, telling us not only countries but also models, should be evaluated on the basis of the deviation ratio and not purely ground truth difference. 
Figure \ref{fig:avg-differences} also serves as a ranking of models' ability to align to cultural values. Therefore, we can conclude that GLM-4 has the strongest ability to align among the models tested. With a closer look, we also note that the aligned languages typically result in better alignment, as with Qwen 7B, despite some exceptions.

\subsection{Model-Origin Analysis}\label{sec:originanalysis}

Now that we have a basic idea of the strongest performing models for cultural alignment and the countries they align with best, we delve deeper into the factors that may be influencing these results. As the development of LLMs is recently being dominated by companies and institutions in two countries, the US and China, the differences in these two countries' models in terms of cultural output are worth investigating. Findings can serve to reinforce or dispel concerns that models are imbued with values from their creators \citep{buyl_large_2024, santurkar_whose_2023}.

\begin{figure}[h!]
    \centering
    \begin{tikzpicture}
        \node[anchor=south west, inner sep=0] (image) at (0,0) 
        {\includegraphics[width=0.72\linewidth]{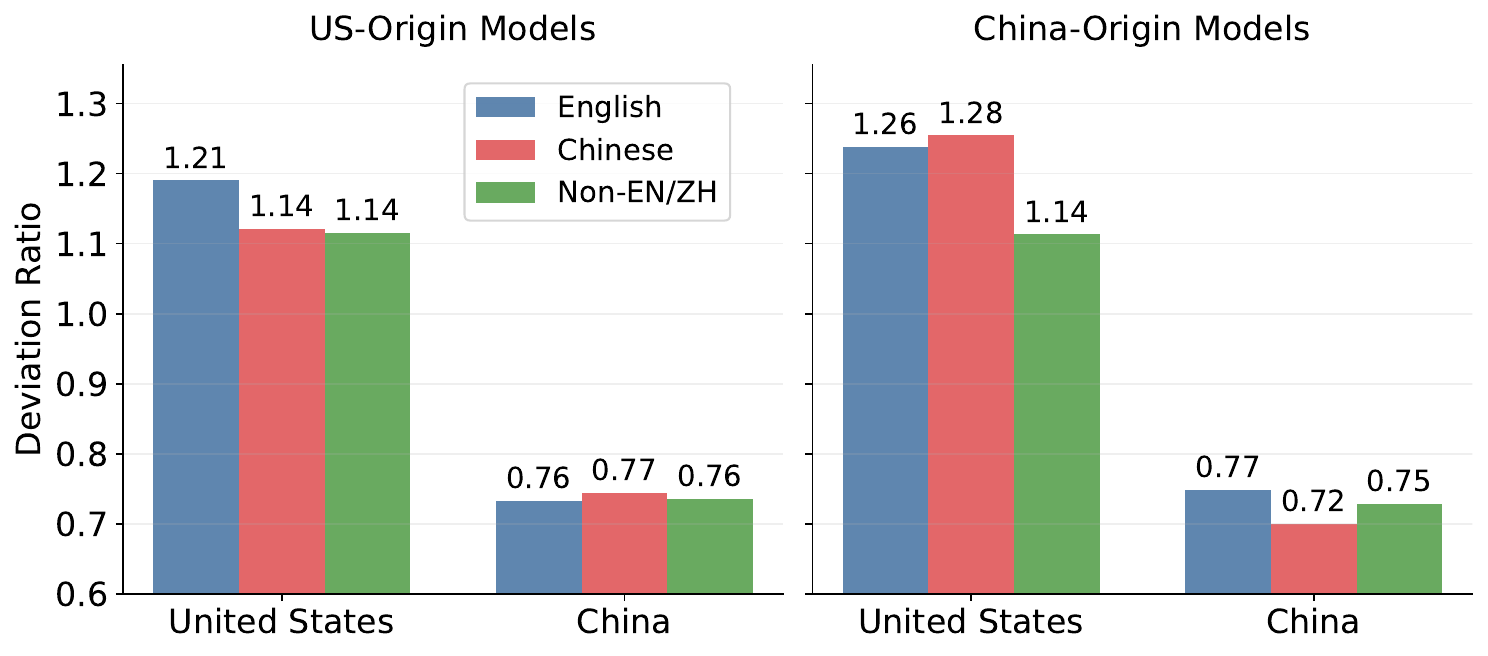}};
        \begin{scope}[x={(image.south east)}, y={(image.north west)}] %
            \node[anchor=north west, text=black] at (0.0, 1) {\textbf{(a)}}; %
            \node[anchor=north west, text=black] at (0.5, 1) {\textbf{(b)}}; %
        \end{scope}
    \end{tikzpicture}
    \caption{Deviation ratio comparison between models of US-origin (a) and China-origin (b), and three forms of prompts: English, Chinese, and the average of all other languages. Scores show alignment to US and China ground truth culture scores. }
    \label{fig:US-China_origin_comparison}
\end{figure}

\begin{figure}
\centering
\includegraphics[width=0.52\linewidth]{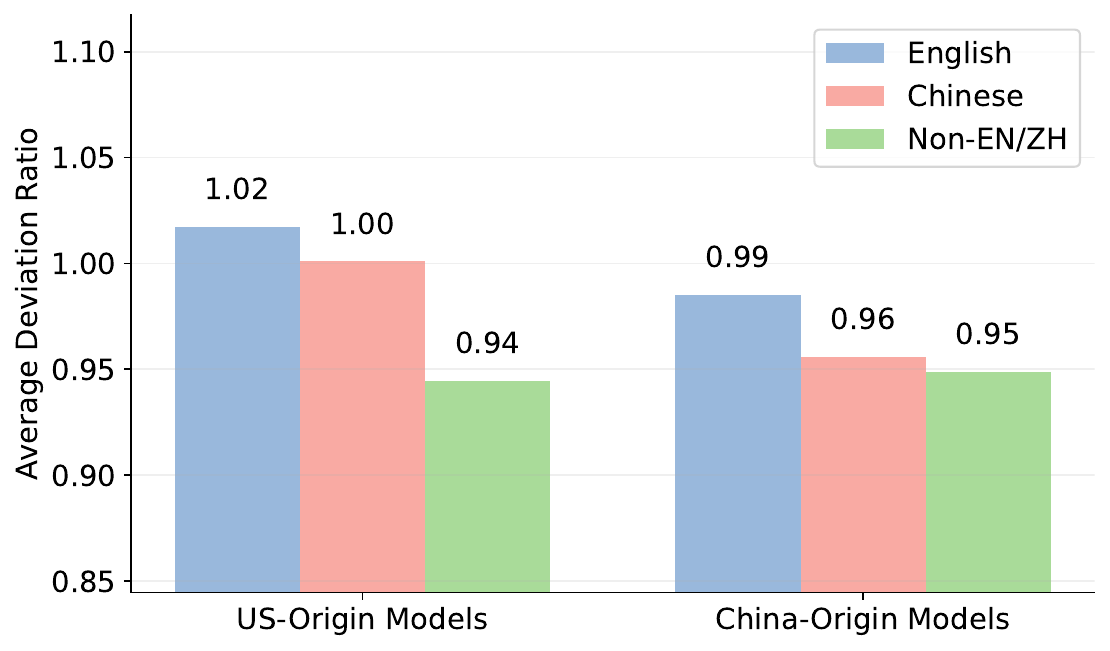 }
\caption{Model origin between US and China-origin models, prompted in English, Chinese, and the average of all other languages. Scores are averaged over all country results to show average alignment w.r.t. model-original and prompt language.}
\label{fig:global_average_origin_comparison}
\end{figure}

First, we aim to determine whether there is an overall difference in alignment between US and China-origin models in exhibiting the cultures of those same two countries in Figure \ref{fig:US-China_origin_comparison}. To make the inquiry more nuanced, we separate results by three forms of language prompting: English, Chinese, and the average of all other languages. For US-origin models, we find that there is an intuitive trend in that English prompting results in better alignment with US culture, and Chinese with China, but there is an opposite trend among China-origin models. We also note that China-origin models are able to align to US culture better than US-origin models, regardless of English or Chinese as the prompt language. 
We also determine the state of alignment on the models of both countries' origins in their culture representations of all countries, illustrated in Figure \ref{fig:global_average_origin_comparison}. We find that on average, US-origin models align better across cultures with prompts in English and Chinese, whereas China-origin models align slightly better using other languages. 

In sum, there is no reason for concern that China-origin models will be imposing any misaligned cultural values on users, as they are able to align very well with US culture in all languages. However, one also might ask why China-origin models are not able to align better with the culture of their origin country. Some possible insights towards this question are provided in the following section.

\subsection{Model-Size and External Factors}

\begin{figure}
    \centering
        \begin{tikzpicture}
        \node[anchor=south west, inner sep=0] (image) at (0,0) {\includegraphics[width=0.95\linewidth]{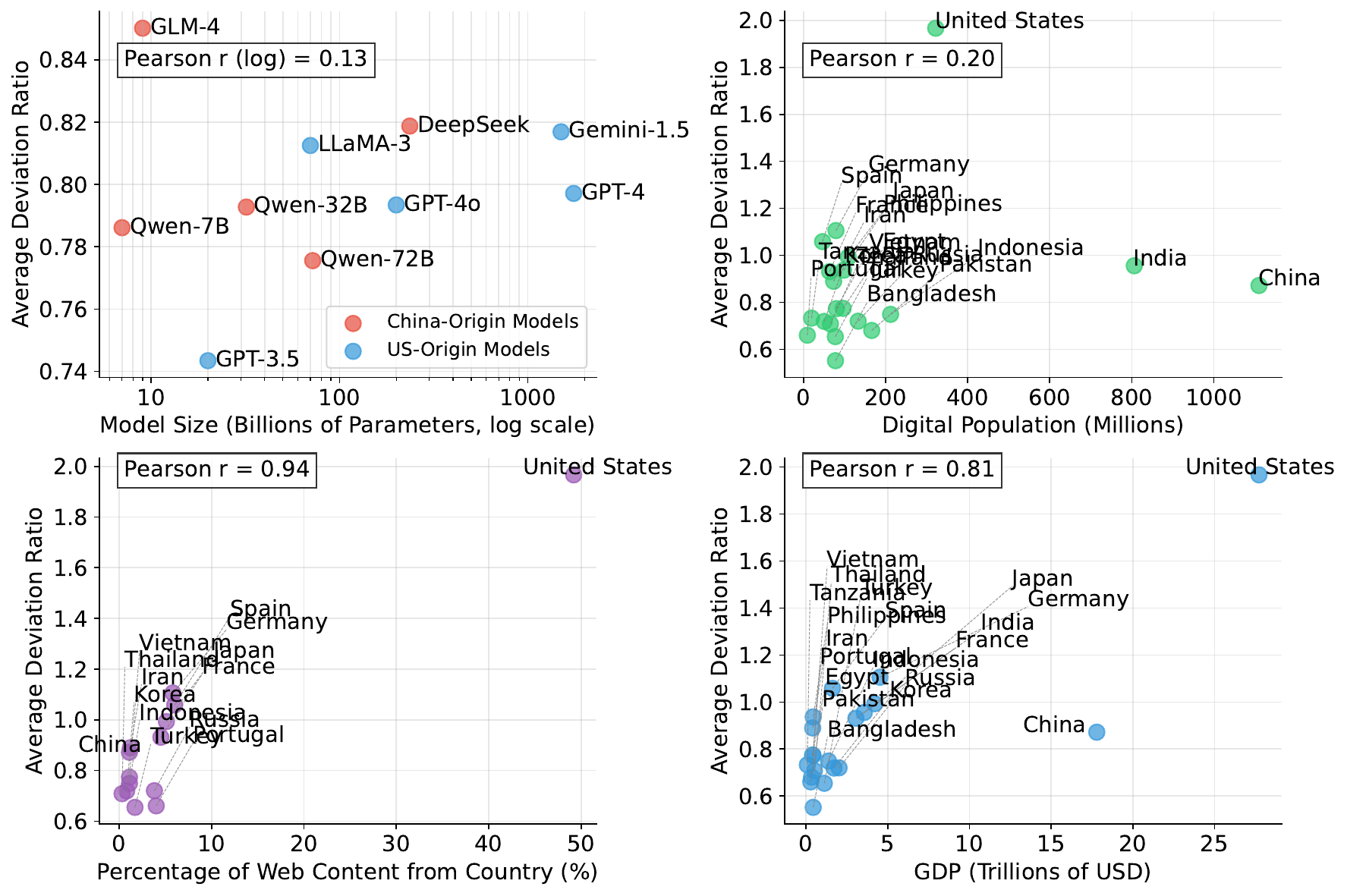}};
        \begin{scope}[x={(image.south east)}, y={(image.north west)}] %
            \node[anchor=north west, text=black] at (-0.02,1.0) {\textbf{(a)}}; %
            \node[anchor=north west, text=black] at (0.5,1.0) {\textbf{(b)}}; %
            \node[anchor=north west, text=black] at (-0.02,0.5) {\textbf{(c)}}; %
            \node[anchor=north west, text=black] at (0.5,0.5) {\textbf{(d)}}; %
        \end{scope}
    \end{tikzpicture}
    
    \caption{Figures comparing alignment with model size and external factors.}
    \label{fig:external_correlations}
\end{figure}

It does not suffice to state which models and countries have the best alignment without asking why this is so. In Figure \ref{fig:external_correlations}, we attempt to answer this question by correlating model size with model alignment and three external factors to country alignment: digital population \footnote{
World Bank, 2023. "Individuals using the Internet (\% of population)", https://data.worldbank.org/indicator/IT.NET.USER.ZS}, percentage of web content \footnote{"Usage statistics of content languages for websites". W3Techs 2024 survey. https://w3techs.com/technologies/overview/content\_language}, and GDP \footnote{World Bank, 2023. "GDP (current US\$)", https://data.worldbank.org/indicator/NY.GDP.MKTP.CD}. For web content per country, we approximate using data for the fraction of content in each country's primary language. 

\begin{figure*}[h]
  \centering
    \begin{subfigure}{0.52 \textwidth}
    \centering
\includegraphics[width=\linewidth, trim=0 0 0 0, clip]{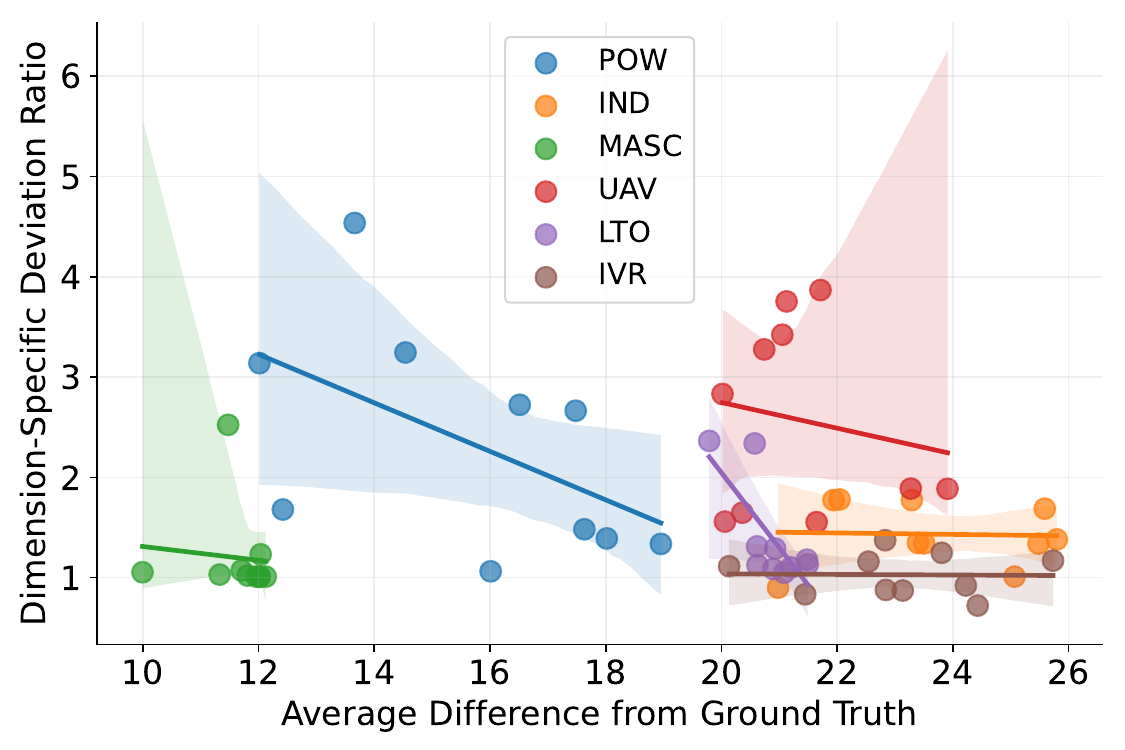} 
  \end{subfigure}
  \vspace{-0.5cm}
  \caption{Ground truth difference vs. dimension-specific deviation ratio (higher DR is better)}
  \label{fig:dimensions}
\end{figure*}

First, in Figure \ref{fig:external_correlations}(a), we see the relationship between model size, given in billions of parameters, and overall alignment. While there is a somewhat log-linear relationship, GLM-4 contradicts expectations by being the strongest aligning model despite having the second lowest number of parameters, only 9 billion. This suggests that there is more to cultural value alignment than just model parameter size. Among Figures \ref{fig:external_correlations}(b) through (d), we note that there are very strong correlations between LLM cultural alignment and two external factors: the quantity of web content per country and national GDP. The main difference between the two is that China's web content proportion is not commensurate with its GDP, causing the person correlation to be lower for Figure \ref{fig:external_correlations}(d) than for (c). However, although the correlation in \ref{fig:external_correlations}(c) is very strong, it is important to note that all countries have both very low cultural alignment scores and content percentages compared to the United States. This means that an important future direction for cultural adaptation in LLMs is to increase the availability of training data from non-US countries. 
Finally, we examine the influence on cultural dimensions and alignment. In Figure \ref{fig:dimensions}, the average deviation ratios over each dimension over all countries and models is displayed. From this figure we observe that certain dimensions are more readily "alignable" than others. Namely, power distance (POW) and uncertainty avoidance (UAV) are more easily aligned to than masculinity (MASC), individualism (IND), and indulgence vs. restraint (IVR). These findings tell us that, perhaps with training data that is more representative of the more difficult dimensions, the obstacle to aligning with more difficult dimensions can be overcome. 
\vspace{0.5cm} %

\vspace{-0.75cm} %

\section{Related Works}\label{sec:related}

Bias in LLMs has garnered wide attention in the research community. \cite{liang2021towards} was one of the first to establish benchmarks and propose mitigating strategies for bias in GPT-2. More recently \cite{tao_auditing_2023} quantify the cultural bias of GPT-3, ChatGPT, and GPT-4 as being close to Western European countries on a two-dimensional cultural scale and mitigates it with a simple cultural prompting strategy. \cite{masoud_cultural_2023} come to similar conclusions showing that LLMs tend to display Western values. 
Notably, \cite{abid2021persistent} find striking evidence of a consistent anti-Muslim bias in LLMs. 
\cite{huang_culturally_2023} find that ChatGPT and GPT-4 tend to normalize culturally interpretable scenarios to an American context based on comparing interpretations between US and Indian annotators. 
Meanwhile, fairness in machine learning has been examined by \cite{barocas2023fairness} and by \cite{blodgett2020language} for the NLP scenario. \cite{jakesch_co-writing_2023} find that writing with an LLM co-writer can cause one's writing to be opinionated in a certain way. In order to help the research community tackle these issues, \cite{dhamala2021bold} provide a dataset of prompts for bias benchmarking. \cite{durmus_towards_2023} also contribute to bias mitigation by designating three types of culture auditing, finding that cross-national prompting succeeds at adapting the model's responses to a specified country's culture. 
In turn, our work is the first to conduct a large-scale, comprehensive evaluation of LLM cultural value alignment over models, countries, and languages,  including previously neglected factors such as model origin
and external factors.

\section{Conclusions}

In this work, we investigate the state of cultural value alignment in LLMs using a wide array of models, countries, and languages to understand which countries are best aligned to, and which models best align to them. We immediately discover that LLMs skew all cultures toward a moderate global average culture that is close to the median possible value for all dimensions. Therefore we devise a metric called the deviation ratio to evaluate alignment in spite of this skew, favoring models that are able to align to countries whose ground truth values are further from the global average. The results of the analysis tell us that the United States culture is most readily aligned by a wide margin, and that this could be in large part explained by the amount of training data available in comparison to other countries. 
Additionally, we find that GLM-4 is able to align best across cultures despite having a very low parameter size, suggesting that there are factors besides model size that are crucial to cultural alignment. We also find that both China-origin models and US-origin models align better to the United States than to China, and that models prompted in English are able to align better to all cultures on average as opposed to Chinese or any other language. 
Some limitations of our study include the countries and languages chosen. There are many countries that have several spoken languages, and since we only test one language per country, our cultural alignment evaluations cannot be considered complete. Furthermore, there is the question of how LLMs align to the cultures of countries that have the same primary language. We leave these tasks for future work. 

\bibliography{colm2025_conference}
\bibliographystyle{colm2025_conference}

\appendix
\section{Appendix}

\renewcommand{\thefigure}{\Alph{figure}}
\renewcommand{\thetable}{\Alph{table}}
\setcounter{figure}{0} %
\setcounter{table}{0} %

\subsection{Cultural Values Survey}\label{sec:appendixhofstede}

The six dimensions of the 24-item questionnaire are Power Distance (POW), Individualism vs. Collectivism (IND), Masculinity vs. Femininity (MASC), Uncertainty Avoidance (UAV), Long-Term Orientation vs. Short-Term Orientation (LTO), and Indulgence vs. Restraint (IVR).
Once responses are obtained for all the items of the survey, culture scores are calculated for each of the six dimensions  using the equation:

\begin{equation}\label{eqn:culturescore}
        \mathcal{D}_i = \lambda_1(Q_1 - Q_2) + \lambda(Q_3 - Q_4) + C_{i},
\end{equation}

\noindent wherein there are four questions associated with each dimension, $\lambda_1$ and $\lambda_2$ represent hyperparameters for calculating each dimension's score provided by the survey creator, and $C_i$ is a constant used to move the scores into a desired range. 

The cultural output values are all normalized to a scale between 0 and 100 to conform to the scale of the ground truth scores. Ground truth scores, in turn, are obtained from the Hofstede official website, where they are aggregated from valid studies~\footnote{https://geerthofstede.com/country-comparison-bar-charts/}, and for any countries whose data was not available, ground truth scores were obtained from a third-party consultancy based upon Hofstede's work~\footnote{https://www.theculturefactor.com/country-comparison-tool}.

To distinguish countries from languages in the text and figures, languages are always given in abbreviated form, whereas country names are written out in full.

\begin{table}[h!]
\centering
\footnotesize
\begin{tabular}{ll} %
\toprule
\textbf{Country} & \textbf{Language (Code)} \\
\midrule
Bangladesh & BEN \\
China & ZH \\
Egypt & AR \\
France & FR \\
Germany & DE \\
India & HIN \\
Indonesia & ID \\
Iran & FA \\
Japan & JA \\
S. Korea & KR \\
Pakistan & UR \\
Philippines & TG \\
Portugal & PT \\
Russia & RU \\
Spain & ES \\
Tanzania & SW \\
Thailand & THA \\
Turkey & TR \\
United States & EN \\
Vietnam & VN \\
\bottomrule
\end{tabular}
\caption{Countries and their corresponding languages }
\label{tab:alphabetical_countries_languages}
\end{table}
\begin{figure}[h!]
    \centering
    \includegraphics[width=0.90\linewidth]{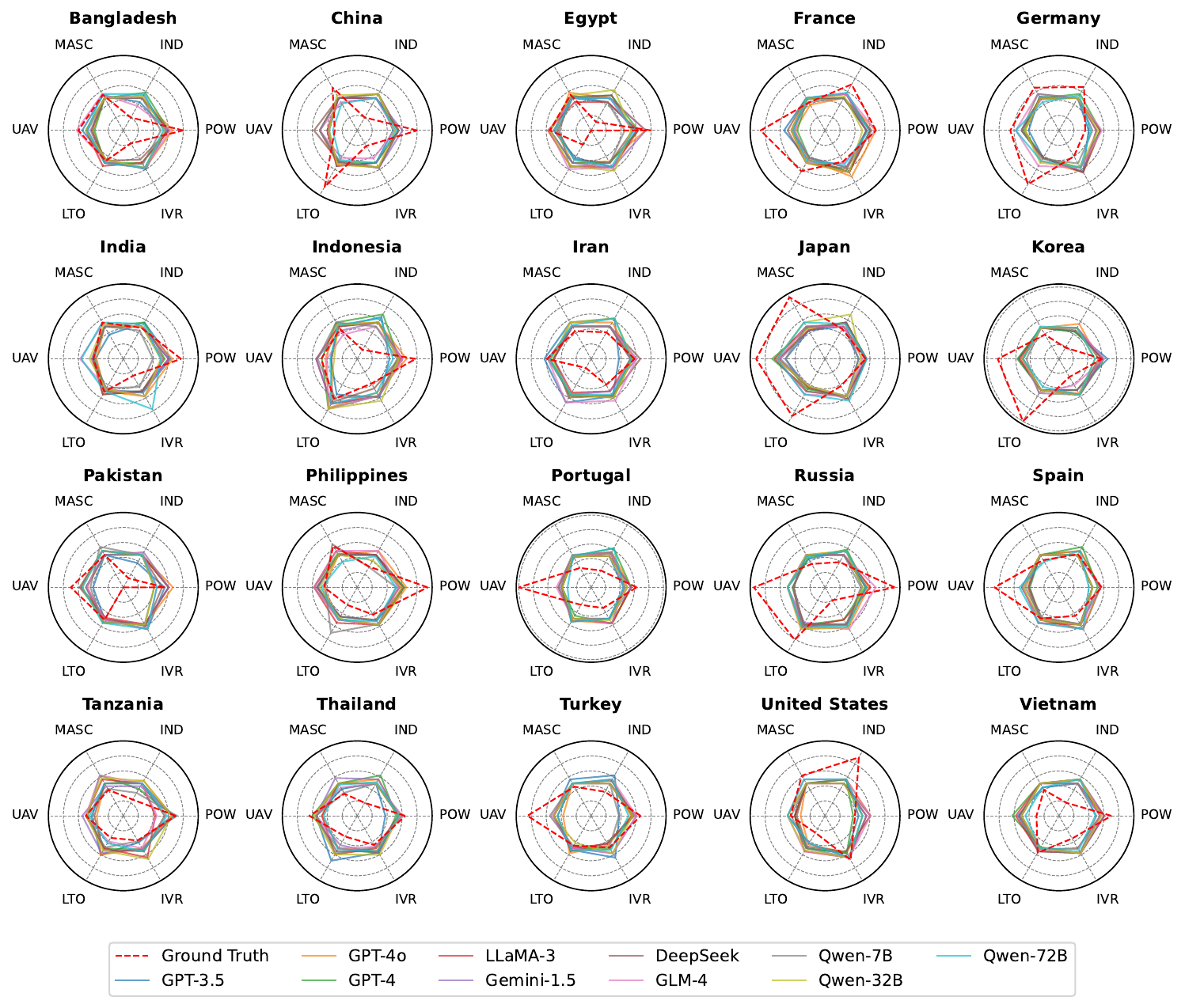}
    \caption{Comparison of raw results for each dimension charts for all countries.}
    \label{fig:country_raw_results_radars}
\end{figure}

\end{document}